\documentclass[prb,twocolumn]{revtex4}
\usepackage{graphicx}
\begin{document}

\title[Waterfall effect in PZN-PT...]{ Origin of the "waterfall"
effect in phonon dispersion of relaxor perovskites}

\date{\today}
\author{J.~Hlinka, S.~Kamba and J.~Petzelt}
\affiliation{Institute of Physics, Academy of Sciences of the
Czech Republic,
   Na Slovance 2, 18221 Praha 8, Czech Republic}
\author{ J.~Kulda}
\affiliation{Institute Laue-Langevin, Avenue des Martyrs, 38640
Grenoble, France}
\author{ C. A. Randall and S. J. Zhang }
\affiliation{Materials Research Institute, Pennsylvania State
University,
  University Park, PA 16802, U.S.A.}

\begin{abstract}
We have undertaken an inelastic neutron scattering study of the
perovskite relaxor ferroelectric Pb(Zn$_{1/3}$Nb$_{2/3}$)O$_3$
with 8\% PbTiO$_3$ (PZN-8\%PT) in order to elucidate the origin of
the previously reported unusual kink on the low frequency
transverse phonon dispersion curve (known as "waterfall" effect).
We show that its position ($q_{\rm wf}$) depends on the choice of
the Brillouin zone and that the relation of $q_{\rm wf}$ to the
size of the polar nanoregions is highly improbable. The waterfall
phenomenon is explained in the framework of a simple model of
coupled damped harmonic oscillators representing the acoustic and
optic phonon branches.
 \end{abstract}

\pacs {77.80.-e, 78.70.Nx, 63.20.Dj, 77.84.Dy}

\maketitle

\section{INTRODUCTION}

Ferroelectric perovskites Pb(Zn$_{1/3}$Nb$_{2/3}$)O$_3$ (PZN) and
Pb(Mg$_{1/3}$Nb$_{2/3}$)O$_3$(PMN) and related materials have
recently attracted a great attention, among others, due to the
excellent piezoelectric properties\cite{Giant} of their solid
solutions with PbTiO$_3$. Their average ABO$_3$ perovskite
structure is perturbed by a short-range occupational ordering on
B-site positions.\cite{Hil90,Ran90} It is known that this
nanoscopic inhomogeneity is responsible for relaxor properties of
these materials, such as smearing and frequency dependence of the
dielectric anomaly associated with the ferroelectric
ordering\cite{Setter}, and for other phenomena reflecting the
presence of polar nanoregions (PNR) persisting hundreds of Kelvins
above the temperature of dielectric permitivity
maximum\cite{Burns}. Unfortunately, it is not easy to deal with
these inhomogeneities in a simple enough model, and the
understanding of the relaxor properties is not yet satisfactory.

Since these materials are available as large single crystals,
inelastic neutron scattering can be used to study the phonon
dispersion curves.
  The spectrum of the lowest frequency transverse
excitations usually consists of interacting and mutually repelling
transverse acoustic and transverse optic phonon branches.
  It is known that upon approaching a displacive
ferroelectric phase transition, the lowest zone center optic mode
(soft mode) tends to zero frequency.
  Instead of that, it was repeatedly found that  near the
"smeared" phase transition of relaxor ferroelectrics, the upper
branch appears to drop precipitously into the lower branch at a
finite value of momentum transfer of the order of $q_{\rm wf} =
0.2\,$\AA$^{-1}$.
  For explanation of this "waterfall" effect,
  Gehring, Park and Shirane\cite{GehPRL00}
proposed that the characteristic wave vector $q_{\rm wf}$
corresponds to the size of the PNR's, and that due to these PNR's
the soft branch phonons with $q< q_{\rm wf}$ cannot effectively
propagate in the crystal.
  The idea was subsequently specified within a mode-coupling
model\cite{GehPRB01,ShiJPSJ} assuming a sharp increase of
transverse optic branch damping (inverse lifetime) for $q < q_{\rm
wf}$.

\begin{figure}[h]
\hspace{0cm} {\includegraphics[angle=270,scale=0.26, clip=true]
 {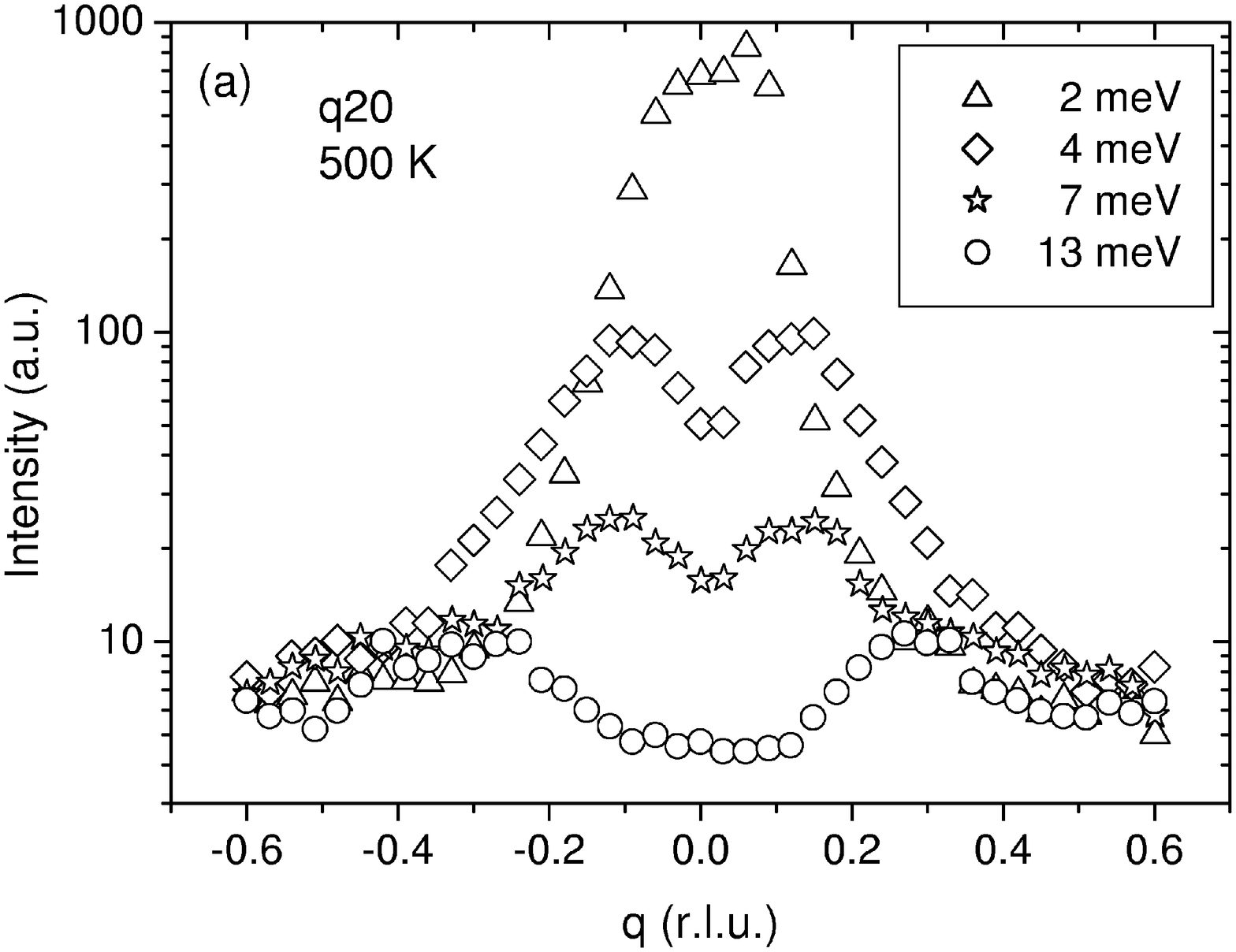}}\\[2mm]
\hspace{1cm} {\includegraphics[angle=270,scale=0.26, clip=true]
 {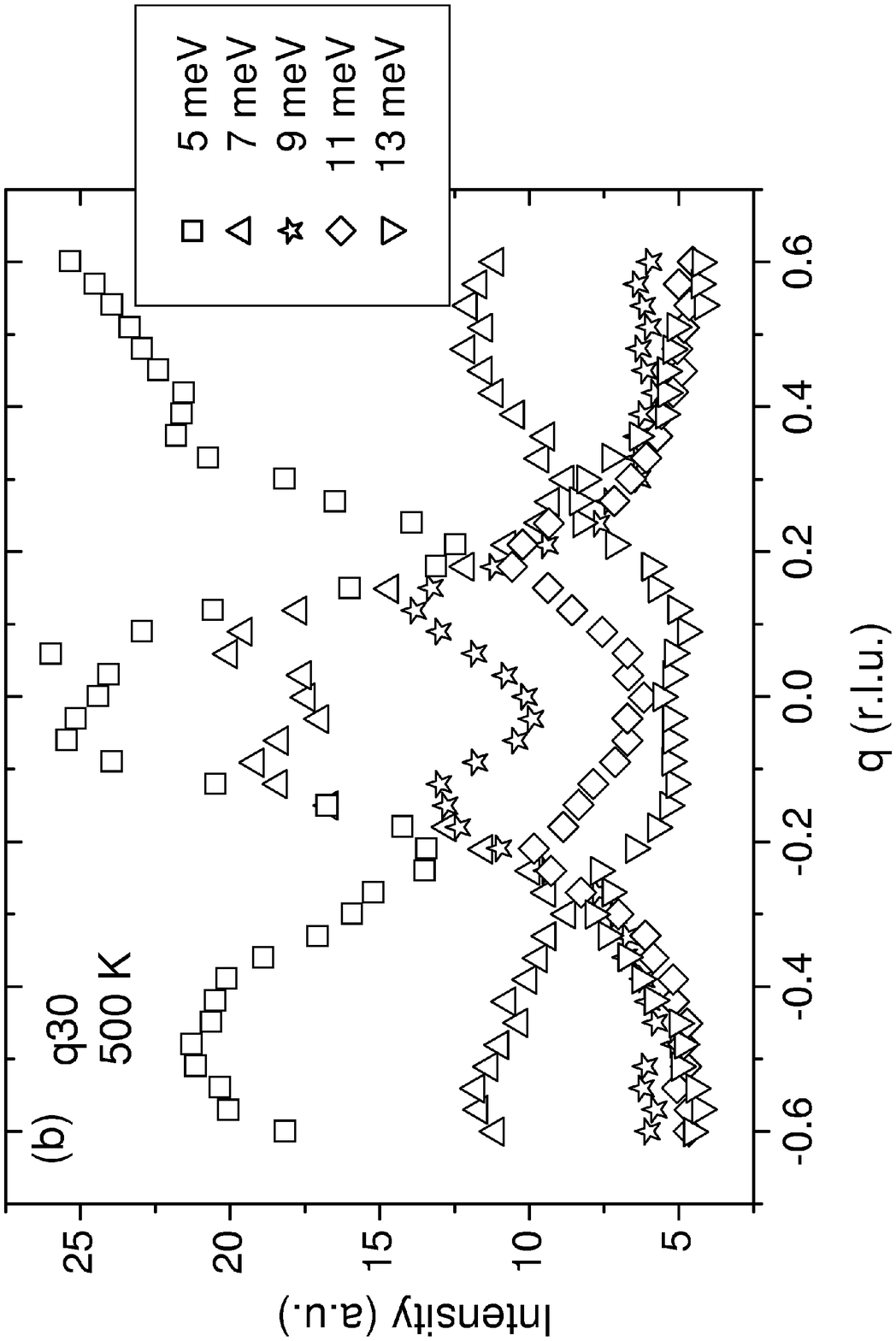}}
\caption{ Low frequency phonon modes in PZN-8\%PT measured by
constant energy scans at 500\,K. Measurements were performed in
(a) 020 Brillouin zone ($Q=(q 2 0) $ scans) and in (b) 030
Brillouin zone ($Q=(q 30) $ scans). \label{Fig1}}
\end{figure}

Within the past three years, it was established that the
"waterfall" effect is common to a number of relaxor
ferroelectrics, including PZN\cite{GehPRB01,Tomeno},
PMN\cite{GehVak00,GehPRL01,Waki02b},
 Pb(Zn$_{1/3}$Nb$_{2/3}$)O$_3$ with
8\% PbTiO$_3$ (PZN-8\%PT)\cite{GehPRL00},
PZN-15\%PT\cite{OraPRB02}, and PMN-20\%PT\cite{Koo02}.
   However, we have noticed that the sharp drop (near $q \approx
0.2\,$\AA$^{-1}$) of the dispersion curves of the single-domain
tetragonal BaTiO$_3$ drawn in Fig.~6 of Ref.~\onlinecite{Shi70}
{\it also} strikingly closely reminds those of relaxors (compare,
for example, with Fig.~2 of Ref.~\onlinecite{GehPRL00}).
    Obviously, the  explanation based on the characteristic size of
PNR's cannot be valid for an experiment performed on a classic
single-domain ferroelectric crystal.
    This letter brings  new inelastic neutron scattering data showing
that such explanation does not hold for relaxors either.
   We demonstrate that the "waterfall" effect can be ascribed to an
entirely classic interference of lineshape anomalies due to the
coupled acoustic and optic branches without any {\it ad-hoc}
assumption of an anomalous increase of the bare optic branch
damping below $q_{\rm wf}$.


For the experiment we have chosen a 3.7 g single crystal of
PZN-8\%PT, used already in our previous neutron study\cite{hli03}.
This as-grown, optically transparent, yellowish single crystal was
produced by the high temperature flux technique at Materials
Research Institute, Pennsylvania State University. This promising
PZN-8\%PT material, showing both relaxor and giant piezoelectric
properties, has been studied
 recently by a number of techniques, and it is the one for which
the possible relation of $q_{wf}$ to the size of PNR's was firstly
invoked.\cite{GehPRL00}

    The present experiment was carried out on the
recently upgraded IN8 thermal neutron spectrometer at the ILL high
flux reactor. The instrument was operated with a fixed wave-number
of scattered neutrons $k_{\rm f} = 3$\,\AA$^{-1}$, using
horizontally focussing (vertically flat) Si crystals as
monochromator and analyzer. The sample was wrapped in a thin Nb
foil and mounted in a vacuum furnace allowing to reach
temperatures up to 1200\,K with a stability better than 1\,K. The
sample was oriented with the cubic [001] axis vertical. This
geometry allowed us to explore the $(hk0)$ scattering plane. Both
transverse and longitudinal momentum resolution widths measured on
the (030) Bragg reflection were better than 0.07 and 0.06 of
reciprocal lattice unit (rlu, where 1\,rlu is $2\pi/a \approx 1.55
$\,\AA$^{-1}$), respectively. This is about equivalent to
resolution widths in a setup with flat PG (pyrolitic graphite)
crystals and 40' Soller collimators used in the previous studies
by other authors. The present setup using elastically bent Si
crystals \cite{KulNIM96} permits to avoid spurious effects from
tails of reflection curves inherent to PG mosaic crystals and
provides an improved energy resolution of 0.8\,meV (FWHM), as
given by the energy profile of elastic incoherent scattering from
our crystal and confirmed by an independent measurement on a
vanadium reference sample.

\begin{figure}[h]
\hspace{0cm} \centerline{\includegraphics[angle=270,width= 7.5cm,
clip=true] {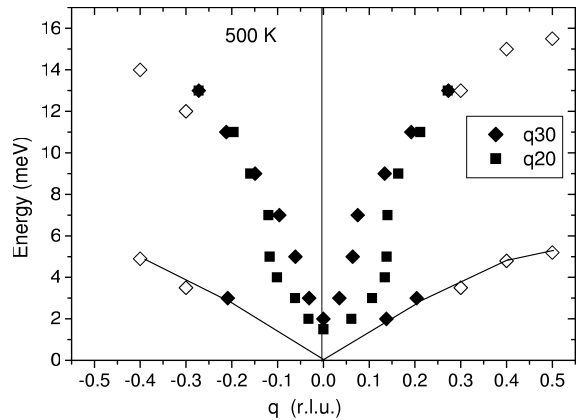}} \caption{ Position of maxima in constant
energy transverse scans performed in 0 2 0 (solid squares) and 0 3
0 (solid diamonds) Brillouin zones (as in Fig. 1). Maxima are
forming two distinct "dispersion curves". The inflection points
near $q_{\rm wf} =0.13$\,rlu and 0.06\,rlu, respectively, are
discussed in the text. Open symbols show maximum intensity
positions in constant-$q$ scans. Lines are only guides for eyes.
  \label{Fig2}}
\end{figure}


     The typical constant energy transverse scans in the 020 and 030 Brillouin
zones at 500\,K are displayed in Fig.~1, and the positions of
their maxima are plotted as solid symbols in Fig.~2. The apparent
dispersion curve obtained in this way from the 020 Brillouin zone
data has an almost vertical rise up near $q_{\rm wf}= 0.13 $\,rlu.
This value corresponds exactly to the $q_{\rm wf}$ obtained from
the 220 zone data at the same temperature and same
crystallographic direction for this material in
Ref.~\onlinecite{GehPRL00}.

   On the other hand, the position of the upward rise observed
   by us in the 030 Brillouin zone is at a significantly smaller
   value of $q_{wf}\approx 0.06$\,rlu.
   As the same [010]-polarized modes with wave vectors
along [100] are probed in both Brillouin zones, the value of
$q_{\rm wf} $ can hardly provide a measure of the size of
PNR's\cite{GehPRL00}.
    On the contrary, the remarkable dependence of $q_{\rm wf} $
    on the chosen Brillouin zone proves that the observed dispersion
is merely an apparent dispersion.
   In fact, it is known that maximum intensity positions for
response function of coupled damped harmonic phonons can be
sensitive to the dynamical structure factors (chosen Brillouin
zone).\cite{Harada}
    Furthermore, the striking difference between the 030 zone
value of $q_{\rm wf} $ and almost the same values of $q_{\rm wf} $
in
 020 and 220 zones indicates that the waterfall effect is
related to dynamical structure factor of bare {\it acoustic}
branch, which is known to be very small in the 030
zone.\cite{Vak02,Waki02b}


As a matter of fact, it turns out that the "waterfall effect" can
be reproduced in a quite simple model. Let us consider the
standard model\cite{Barker,Currat,phason} of two coupled damped
harmonic oscillators, defined by a $2 \times 2$ dynamical and
damping matrices ${\bf D}_q$ and ${\bf \Gamma}_q$
\begin{eqnarray}
{\bf D}_q& =& \left( \begin{array}{cc} \omega^2_{\rm TA}(q)& \Delta (q)\\
 \Delta (q)^* &\omega^2_{\rm TO}(q)
\end{array} \right)\, ,\\
{\bf \Gamma_q} &= &\left( \begin{array}{cc} \Gamma_{\rm TA}(q)& \Gamma_{\rm AO}(q)\\
\Gamma_{\rm AO}(q) &\Gamma_{\rm TO}(q)
\end{array} \right)\, ,
\end{eqnarray}
where $\omega_{\rm TA}(q)$ and $\omega_{\rm TO}(q)$ describes
dispersion of bare acoustic and optic branches, $\Delta (q)$
describes their mutual bilinear interaction, $\Gamma_{\rm TA}(q)$
and $\Gamma_{\rm TO}(q)$ stands for bare mode frequency
independent damping, $\Gamma_{\rm AO}(q)$ stands for viscous
interaction (bilinear in time derivatives of bare mode
coordinates) and $q$ is the reduced phonon wave vector in a chosen
direction (in this letter, along the cubic [100]). In the high
temperature limit ($h\omega << kT $), the corresponding inelastic
neutron scattering intensity is proportional
to\cite{Harada,Currat}
\begin{equation}
I(\omega,q)= kT ~ \omega^{-1}~ {\bf f}(q)^*~ .~{\rm Im} \left[
{\bf G}(\omega,q)
 \right] ~.~ {\bf
f}(q)~,
\end{equation}
\begin{equation}
{\bf G}(\omega,q)= ( {\bf D}_q - i. \omega {\bf \Gamma}_q -
\omega^2 \rm {\bf E})^{-1}\, ,
\end{equation}

where {\bf E} is a $2 \times 2$ unit matrix and {\bf f}(q) is a
2-component vector composed of dynamical structure factors of bare
acoustic and optic branches, respectively. At small frequencies,
it is expected that $\omega . |\Gamma_{\rm AO}(q)| <<
|\Delta_{q}|$ so that, as e.g. in the Ref.~\onlinecite{Harada},
$\Gamma_{\rm AO}(q)=0$ will assumed in the following. The wave
vector dependence of dynamical matrix ${\bf D}_q$ can be
conveniently cast in the form which appears in the simplest
nearest-neighbor interaction models with two degrees of freedom
per unit cell\cite{Etxe,hlinka,Chen}:
\begin{eqnarray}
\omega^2_{\rm TA}(q) &= &A \sin^2( \pi q/2) \, ,\\
\omega^2_{\rm TO}(q) &= & c+B\sin^2( \pi q/2) \, ,\\
\Delta_{q}&= & d \sin^2( \pi q/2)\, ,
\end{eqnarray}
where the real parameters $A, B, c$ and $d$ define the strength of
dispersion of bare acoustic and optic branches, the square of soft
mode frequency and the strength of mode mixing at zone boundary,
respectively.
   Let us note that the dynamical matrix parametrized
in this way shows the same asymptotic behavior for $q \rightarrow
0$ as the standard model of Ref.~\onlinecite{Axe}, and at the same
time provides the correct dispersion behavior also at the zone
boundary. In the same spirit, we introduce
\begin{eqnarray}
\Gamma_{\rm TA}(q) &= & g \sin^2( \pi q/2) \, ,\\
\Gamma_{\rm TO}(q) &= & h,
\end{eqnarray}
where the real parameters $g$ and $h$ define the damping of the
bare acoustic branch (with required asymptotic $q^2$ dependence)
and the q-{ \em independent} damping of bare optic mode,
respectively.

Numerical calculations were performed for a set of realistic
values ($A = 100$\,meV$^2$, $B = 150$\,meV$^2$, $c = 15$\,meV$^2$,
$d = 100$\,meV$^2$, $g = 6$\,meV, $h = 6$\,meV). Constant energy
profiles calculated for the case of equal bare acoustic and optic
structure factors (${\bf f}= (0.5,0.5)$) are shown in Fig.~3,
while phonon dispersion curves obtained by diagonalization of the
dynamical matrix ${\bf D}_q$ are shown in Fig. 4. Fig. 4 also
shows that the positions of the maximum intensity in constant
energy profiles shown in Fig. 3 forms an {\em apparent dispersion
branch} which follows the acoustic branch for small $q$, but then
it sharply rises towards the upper phonon branch near 0.1\,rlu.
The shape of this apparent dispersion curve indeed corresponds
well with the experimental findings shown for example in Fig. 2.

\begin{figure}[h]
\hspace{0cm} \centerline{\includegraphics[angle=270,width= 7.5cm,
clip=true]
 {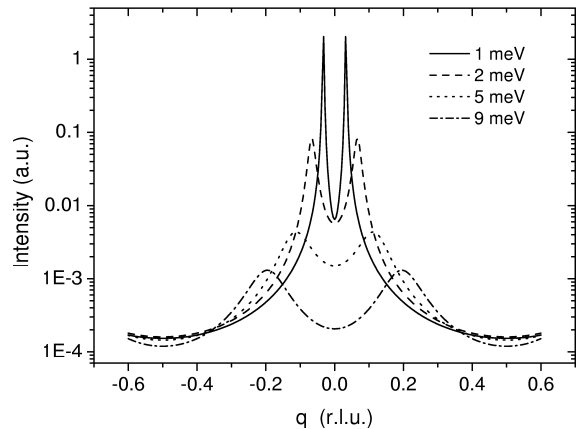}}
\caption{ Constant energy scans at 1-9 meV simulated from the
model described in the text, for structure factors ${\bf f}=
(0.5,0.5)$. \label{Fig3}}
\end{figure}

\begin{figure}[h]
 \hspace{0cm} \centerline{\includegraphics[angle=270,width= 7cm, clip=true]
 {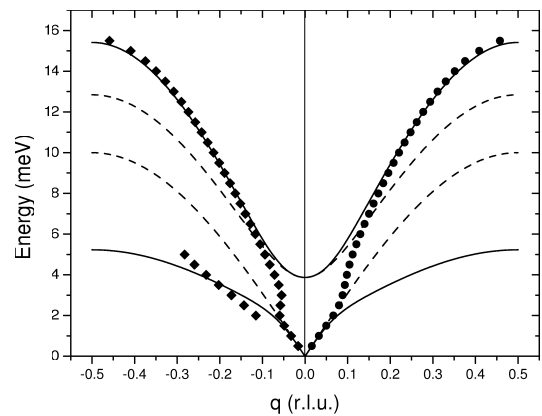}}
\caption{ Dispersion curves calculated from the model described in
the text. Full and dashed lines correspond to frequencies of
repelled and bare phonon branches obtained as square roots of
eigenvalues and of diagonal elements of the dynamical matrix ${\bf
D}_q$, respectively. Full squares (on the right side) stands for
positions of maximum intensity in constant energy profiles shown
in Fig. 3, calculated for balanced structure factors ${\bf f}=
(0.5,0.5)$. Full diamonds (left side of diagram) are calculated in
a same way for the case of a {\em smaller bare acoustic structure
factor} (${\bf f}= (0.25,0.75)$). \label{Fig4}}
\end{figure}

As expected, the "waterfall" region of the apparent dispersion
curve depends noticeably on the dynamical structure factor. For
example, a decrease of the weight of the bare acoustic structure
factor shifts the "waterfall" position towards the zone center, as
it follows from Fig. 4. A similar tendency is found in the
experimental data taken in the "predominantly optic" 030 Brillouin
zone (Fig 2). In the limit of zero bare acoustic structure factor
(${\bf f}= (0,1)$), the waterfall phenomenon completely disappears
(Fig. 5). On the other hand, with a slight change of parameters,
the apparent waterfall dispersion curve can attain even a negative
slope. These results show that constant energy scan technique for
investigation of coupled acoustic-optic branches should be used
with certain caution. More detailed discussion of the model
properties and quantitative analysis of the experimental data is
beyond the scope of this letter and will be published elsewhere.

\begin{figure}[h]
 \hspace{0cm} \centerline{\includegraphics[angle=270,width= 7cm, clip=true]
{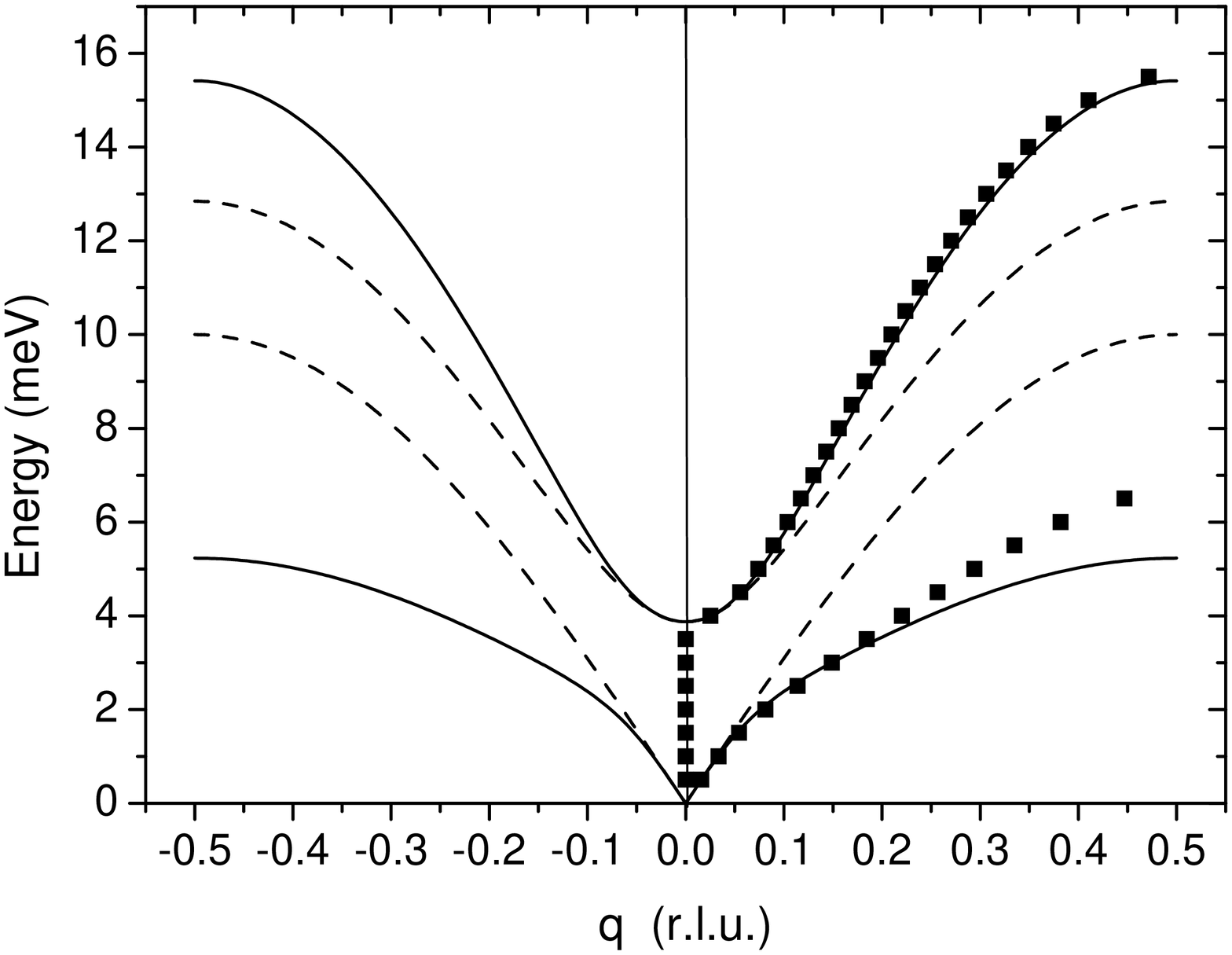}} \caption{ Dispersion curves calculated from the model
described in the text. Symbols have the same meaning as in Fig 4.;
full squares are calculated from constant energy profiles
calculated for the case of {\em zero bare acoustic structure
factor} (${\bf f}= (0,1)$). \label{Fig5}}
\end{figure}

 In conclusion, we argue that the "waterfall" kink seen on the
 dispersion curves of several relaxor perovskites can be
  explained in the framework of   coupled damped harmonic
 oscillator models describing the interaction of the
 acoustic and optic branches.
 This kink  only appears on an apparent dispersion curve, obtained from
 maxima of constant energy sections of scattered intensity, while
 no such kink is present in the dispersion
 of bare or coupled branches themselves.
 In the light of these results, there is no reason for the position
 $q_{\rm wf}$ of the waterfall kink to be related to the size of the
  polar nanoregions. First of all,
  there is no need to assume the
 unusual increase of bare optic branch damping below the waterfall
  wave vector $q_{\rm wf}$ in the present model. Secondly,
 the "waterfall" position is very sensitive to the dynamical
structure factor both in our model and in our experimental data,
where it becomes reduced by as much as a factor of 2 when passing
from the 020 to 030 Brillouin zone. It also follows that the
"waterfall" anomaly in the low-energy phonon dispersion is not
specific to relaxor ferroelectrics, but it can be encountered in
other systems with interacting acoustic and soft optic branches,
such as the ferroelectric perovskite BaTiO$_3$.


\begin{acknowledgements} The work has been
  supported by Czech grants (projects  A1010213, 202/01/0612, M\v{S}MT LA043
  and AVOZ1-010-914).
\end{acknowledgements}


\begin{thebibliography} {99}
\bibitem{Giant} S. E. Park and  T. R. Shrout,   J. Appl. Phys. {\bf
82}, 1804 (1997)
\bibitem{Hil90}  A. D. Hilton, D. J. Barber, C. A.  Randall, and T. R.
Shrout, J. Mat. Sci. {\bf 25}, 3461 (1990)
\bibitem{Ran90} C. A. Randall and  A. S. Bhala,
 Jpn. J. Appl. Phys. {\bf 29}, 327 (1990)
\bibitem{Setter} N. Setter and L. E. Cross, J. Appl. Phys. {\bf 51},
4356 (1980); F. Chu, N. Setter and A. K. Tagantsev, J. Appl. Phys.
{\bf 74}, 5129 (1993)
\bibitem{Burns} G. Burns  and F. H. Dacol,  Ferroelectrics {\bf 52}, 103 (1983);
Sol. State Comm. {\bf 48}, 853 (1983)
\bibitem{GehPRL00} P. M. Gehring, S. E. Park, and  G. Shirane,
Phys. Rev. Lett. {\bf 84}, 5216 (2000)
\bibitem{GehPRB01} P. M. Gehring, S. E. Park, and  G. Shirane,
Phys. Rev. B {\bf 63}, 224109 (2001)
\bibitem{ShiJPSJ} G. Shirane and P. M. Gehring,
J. Phys. Soc. Jpn. {\bf 70}, 227 (2001)
\bibitem{Tomeno} I. Tomeno, S. Shimanuki, Y. Tsunoda, and Y. Y.
Ishii, J. Phys. Soc. Jpn. {\bf 70}, 1444 (2001)
\bibitem{GehVak00} P. M. Gehring, S. B. Vakhrushev, and G.
Shirane: {\it Fundamental Physics of Ferroelectrics 2000: Aspen
Center for Physics Winter Workshop}, editor R.\ E.\ Cohen, AIP
Conference Proceedings, Melville, New York {\bf 535}, 314 (2000)
\bibitem{Waki02b} S. Wakimoto, C. Stock, Z.-G Ye, W. Chen,  P. M. Gehring,
and G. Shirane, Phys. Rev. B {\bf 66}, 224102 (2002)
\bibitem{GehPRL01} P. M. Gehring,  S. Wakimoto, Z.-G Ye,
and G. Shirane, Phys. Rev. Lett. {\bf 87}, 277601 (2001)
\bibitem{OraPRB02} D. La-Orauttapong, B. Noheda, Z.-G. Ye,  P. M. Gehring,
 J. Toulouse, D. E.  Cox, and  G. Shirane,
 Phys. Rev. B {\bf 65}, 144101 (2002)
\bibitem{Koo02}T. Y. Koo, P. M. Gehring, G. Shirane, V. Kiryukhin, S. G. Lee,
and S. W. Cheong, Phys. Rev. B {\bf 65}, 144113 (2002)
\bibitem{Shi70} G. Shirane, J. D. Axe, and J. Harada, Phys. Rev. B {\bf 2}, 3651 (1970)
\bibitem{hli03} J. Hlinka, S.~Kamba, J.~Petzelt, J.~Kulda, C. A.
Randall, and S. J. Zhang, J. Phys. Cond. Mat. {\bf 15}, 4249
(2003)
\bibitem{KulNIM96} J. Kulda and J. \v{S}aroun, Nucl. Inst. Meth. A {\bf 379},
155 (1996)
\bibitem{Harada} J. Harada, J. D. Axe, and G. Shirane,
 Phys. Rev. B {\bf 4}, 155 (1971)
\bibitem{Vak02} S. B. Vakhrushev and S. M. Shapiro,
Phys. Rev. B {\bf 66}, 214101 (2002)
\bibitem{Barker} A. S. Barker and J. J. Hopfield,
 Phys. Rev. {\bf 135}, A1732 (1964)
\bibitem{Currat} R. Currat, H. Buhay, C. H. Perry and A. M.
Quittet, Phys. Rev. B {\bf 40}, 10741 (1989)
\bibitem{phason} J. Hlinka, I. Gregora and V. Vorl\'{\i}\v{c}ek.
Phys. Rev B {\bf 56}, 13855 (1997)
\bibitem{Etxe} I. Etxebarria, M. Quilichini, J. M. Perez-Mato, P. Boutrouille,
 F. J. Zuniga, and T. Breczewski,  J. Phys. Cond. Mat. {\bf 4},
 8551 (1992)
\bibitem{hlinka} J. Hlinka, M. Quilichini, R. Currat, and J. F. Legrand,
   J. Phys.: Condens. Matter {\bf 8}, 8221  (1996)
\bibitem{Chen} Z. Y. Chen and M. B. Walker,
Phys. Rev. B {\bf 43}, 5634 (1991)
\bibitem{Axe} J. D. Axe, J.
Harada, and G. Shirane, Phys. Rev. B {\bf 1}, 1227 (1969)


\end{thebibliography}
\end{document}